\documentclass[twocolumn,aps,superscriptaddress,floatfix]{revtex4}
\usepackage[latin1]{inputenc}
\usepackage{amsmath}
\usepackage[english]{babel}
\usepackage{amsfonts}
\usepackage{amssymb}
\usepackage{mathtools}
\usepackage{graphicx}
\usepackage[bookmarks]{hyperref}
\usepackage{dsfont}
\usepackage{dcolumn}
\newcommand{\kk}{\underline{k}}
\newcommand{\yy}{\underline{y}}
\newcommand{\wt}[1]{\widetilde{#1}}
\newcommand{\deriv}[2]{\frac{d\, #1}{d #2}}
\newcommand{\parderiv}[2]{\frac{\partial\, #1}{\partial #2}}
\newcommand{\funcint}[1]{\int\mathcal{D}[#1]}

\newcommand{\expval}[1]{\left\langle #1\right\rangle}

\begin{document}

\title{\lq Sinc\rq{}-Noise for the KPZ Equation}

\author{Oliver Niggemann}
\affiliation{Universit\"at W\"urzburg, Fakult\"at f\"ur  Physik und Astronomie, 97074  W\"urzburg, Germany,}
\author{Haye Hinrichsen}
\affiliation{Universit\"at W\"urzburg, Fakult\"at f\"ur  Physik und Astronomie, 97074  W\"urzburg, Germany,}

\begin{abstract}
In this paper we study the one-dimensional Kardar-Parisi-Zhang equation (KPZ) with correlated noise by field-theoretic dynamic renormalization group techniques (DRG). We focus on spatially correlated noise where the correlations are characterized by a \lq sinc\rq{}-profile in Fourier-space with a certain correlation length $\xi$. The influence of this correlation length on the dynamics of the KPZ equation is analyzed. It is found that its large-scale behavior is controlled by the \lq standard\rq{} KPZ fixed point, i.e. in this limit the KPZ system forced by \lq sinc\rq{}-noise with arbitrarily large but finite correlation length $\xi$ behaves as if it were excited with pure white noise. A similar result has been found by Mathey et al. \href{https://journals.aps.org/pre/abstract/10.1103/PhysRevE.95.032117}{[Phys.Rev.E 95, 032117]} in 2017 for a spatial noise correlation of Gaussian type ($\sim e^{-x^2/(2\xi^2)}$), using a different method. These two findings together suggest that the KPZ dynamics is universal with respect to the exact noise structure, provided the noise correlation length $\xi$ is finite.
\end{abstract}

\maketitle
\parskip 1mm

\section{Introduction}\label{Introduction}

The \lq standard\rq{} form of the KPZ equation introduced in 1986 by Kardar, Parisi and Zhang, for modeling non-linear growth processes, reads
\begin{align}
 \parderiv{h(x,t)}{t}=\nu\,\nabla^2h(x,t)+\frac{\lambda}{2}\,\left(\nabla h(x,t)\right)^2+\eta(x,t),
 \label{KPZ_Equation}
\end{align}
where $h(x,t)$ is a scalar height field (with $x,\,t$ as space and time coordinates, respectively), $\nu$ is a surface tension parameter and $\lambda$ is a non-linear coupling constant. Here $\eta(x,t)$ denotes an uncorrelated Gaussian noise with zero mean (white noise in space and time) \cite{Kardar_Parisi_Zhang1986}. Therefore the first and second moments of the Gaussian noise are given by
\begin{align}
\begin{split}
\expval{\eta(x,t)}&=0,\\
\expval{\eta(x,t)\eta(x^\prime,t^\prime)}&=2\,D\,\delta^d(x-x^\prime)\delta(t-t^\prime), 
\end{split}
\label{White_Noise}
\end{align}
where $D$ is a constant amplitude. Besides the noise correlation of \eqref{White_Noise}, various spatially and temporally correlated driving forces have been studied over the years \cite{Medina1989,Forster_Nelson_Stephen1977}.\par
With respect to spatial correlations a widely studied type of driving forces is power-law correlated noise with Fourier-space correlations $\sim q^{-2\rho}$ and $\rho>0$ as a free parameter \cite{Medina1989,Janssen1999,Canet2014}. The intriguing observation here is the emergence of a new \lq noise\rq{} fixed point for $\rho>1/4$, additionally to the standard Gaussian and KPZ fixed points.\par
Recently the KPZ equation with spatially colored and temporally white noise decaying as $\sim e^{-x^2/(2\xi^2)}$ was studied by a non-perturbative DRG analysis in \cite{Canet2017}. It was found that for small values of $\xi$ the KPZ equation behaves in the large-scale limit as if it were stirred by white noise, i.e. with a driving force with vanishing correlation length. Since the non-perturbative RG equations are difficult to solve analytically, the authors of \cite{Canet2017} relied on numerical techniques.\par
In the present paper we study the case of spatially correlated noise, where the correlations are characterized by a \lq sinc\rq{}-like profile in Fourier-space. As in \cite{Canet2017}, these correlations are characterized by a finite correlation length $\xi$. Unlike \cite{Canet2017}, we solve the problem analytically by using field-theoretic DRG techniques.\par
For treating \lq sinc\rq{}-type noise, we first generalize the field-theoretic DRG formalism for the KPZ equation in such a way that we can handle homogeneous and isotropic noise distributions, whose correlations in momentum space are given by
\begin{align}
 \expval{\eta(q,\omega)\eta(q^\prime,\omega^\prime)}=2\,D(|q|^2)\,\delta^d(q+q^\prime)\,\delta(\omega+\omega^\prime).
 \label{General_Noise_Classification}
\end{align}
Note that $D$ does not depend on the frequency, i.e., the noise is spatially colored but temporally white.\par
For this class of noise correlations, which includes the power-law ($\sim q^{-2\rho}$), Gaussian ($\sim e^{-\xi^2 q^2/2}$) and \lq sinc\rq{}-type correlations, the field theoretic DRG formalism will be built in the next section. With the theoretical framework laid out in \autoref{Generalized Field Theoretic Renormalization Group Procedure}, the explicit \lq sinc\rq{}-noise excitation will be analyzed in \autoref{The KPZ Equation with Sinc-Noise Correlation}. In \autoref{Discussion} the results obtained in \autoref{The KPZ Equation with Sinc-Noise Correlation} will be discussed.

\section{Generalized Field Theoretic Renormalization Group Procedure}\label{Generalized Field Theoretic Renormalization Group Procedure}

A useful tool for building a field theory for stochastic differential equations of type \eqref{KPZ_Equation} is the effective action $\mathcal{A}[\wt{h},h]$, known as the Janssen-De Dominicis response functional \cite{Janssen1976,DeDominicis1976}. Here the action depends on the original height field $h(x,t)$ and the Martin-Siggia-Rose (MSR) auxiliary field $\wt{h}(x,t)$.\par
To derive the effective action, it is useful to transform \eqref{General_Noise_Classification} into real-space:
\begin{align}
 \expval{\eta(x,t)\eta(x^\prime,t^\prime)}=2D(x-x^\prime)\delta(t-t^\prime),
 \label{General_Noise_Correlation_Real_Space}
\end{align}
where $D(x)=\mathcal{FT}^{-1}\{D(|q|^2)\}$ \cite{Janssen1999,Canet2014}.\par
Using the abbreviations $\yy=(x,t)$ and $\int_{\yy}\cdots=\int d^dx\int dt\cdots$, the corresponding Gaussian noise probability distribution can be written as \cite{Hochberg1999}
\begin{align}
 \mathcal{W}[\eta]\propto\exp\left[-\frac{1}{2}\,\int_{\yy}\int_{\yy^\prime}\,\eta(\yy)\mathcal{M}(\yy;\yy^\prime)\eta(\yy^\prime)\right],
 \label{Noise_Distribution_General}
\end{align}
where $\mathcal{M}(x,t;x^\prime,t^\prime)$ is the inverse of the \lq covariance operator\rq{} 
\begin{align*}
 \mathcal{M}^{-1}(x,t;x^\prime,t^\prime)=\expval{\eta(x,t)\eta(x^\prime,t^\prime)} 
\end{align*}
given in \eqref{General_Noise_Correlation_Real_Space}, i.e.
\begin{align}
\begin{split}
 &\int d^dx^\prime\int dt^\prime\,\mathcal{M}(x,t;x^\prime,t^\prime)\mathcal{M}^{-1}(y,\tau;x^\prime,t^\prime)\\
 &\qquad=\delta^d(x-y)\delta(t-\tau).
 \end{split}
 \label{Inverse_Identity}
\end{align}
Using \eqref{Noise_Distribution_General} and following Refs \cite{Canet2010,Munoz1989,Hochberg2000}, the expectation value of any observable $O[h]$ can be written as
\begin{widetext}
 \begin{align}
  \begin{split}
 \expval{O[h]}&=\funcint{h}\funcint{i\wt{h}}O[h]\exp\left[\int d^dx\int dt\,\wt{h}(x,t)\left(\partial_th(x,t)-\nu\nabla^2h(x,t)-\frac{\lambda}{2}\left(\nabla h(x,t)\right)^2\right)\right]\times\\
 &\qquad\qquad\times\funcint{\eta}\exp\left[-\frac{1}{2}\,\int d^dx\int dt\left(\int d^dx^\prime\int dt^\prime\,\eta(x,t)\mathcal{M}(x,t;x^\prime,t^\prime)\eta(x^\prime,t^\prime)-\wt{h}(x,t)\eta(x,t)\right)\right].
 \end{split}
 \label{Exp_Val_Noise_History}
 \end{align}
\end{widetext}
Eq. \eqref{Exp_Val_Noise_History} can be rewritten in the form \cite{Janssen1999,Canet2014,Canet2016,Janssen1976,DeDominicis1976}
\begin{align}
\begin{split}
 \expval{O[h]}&\propto\funcint{h}\,O[h]\mathcal{P}[h]\\
 &=\funcint{h}\,O[h]\funcint{i\wt{h}}e^{-\mathcal{A}[\wt{h},h]},
 \label{Exp_Val_JD_Func}
 \end{split}
\end{align}
with the Janssen-De Dominicis functional \cite{Taeuber2014_Book,Hochberg1999,Hochberg2000,Canet2011_1}
\begin{align}
\begin{split}
 \mathcal{A}&[\wt{h}(x,t),h(x,t)]\\
 &=\int_{\yy}\left\lbrace\wt{h}(\yy)\left(\parderiv{h(\yy)}{t}-\nu\nabla^2h(\yy)-\frac{\lambda}{2}\left(\nabla h(\yy)\right)^2\right)\right.\\
 &\left.\quad-\int d^dx^\prime\,\wt{h}(x,t)D(x-x^\prime)\wt{h}(x^\prime,t)\right\rbrace.
 \end{split}
  \label{Janssen_De_Dominicis_Real_Space}
\end{align}
With this functional, one can carry out the usual field-theoretic perturbation expansion in $\lambda$, see e.g. \cite{Taeuber1994,Taeuber2014_Book,Zinn_Justin2002,Zinn_Justin2007}.\par
The KPZ equation is known to be invariant under tilts (Galilei transformation) of the form \cite{Taeuber1994,Janssen1999}
\begin{align}
 \begin{split}
  h(x,t)&\to h^\prime(x,t)=h(x+\alpha\,\lambda\,t,t)+\alpha\cdot x,\\
  \wt{h}(x,t)&\to\wt{h}^\prime(x,t)=\wt{h}(x+\alpha\,\lambda\,t,t),
 \end{split}
 \label{Tilting_Symmetry}
\end{align}
where $\alpha$ is the tilting angle. This symmetry is giving rise to two Ward-Takahashi identities. For this reason the KPZ equation has only two independent RG parameters, namely, the noise correlation amplitude $D(x-x^\prime)$ and the surface tension $\nu$ \cite{Taeuber1994,Janssen1999}.\par
These parameters are renormalized by
\begin{align}
\begin{split}
 D_R&=Z_DD,\\
 \nu_R&=Z_\nu\nu,
 \end{split}
 \label{D_R_Nu_R}
\end{align}
where the multiplicative RG factors $Z_\nu$ and $Z_D$ compensate logarithmic UV divergences occurring in the perturbation integrals. They are related to $\Gamma_{\wt{h}h}$ and $\Gamma_{\wt{h}\wt{h}}$, respectively.
\begin{figure*}[!t]
\centering
 \includegraphics[width=0.9\linewidth]{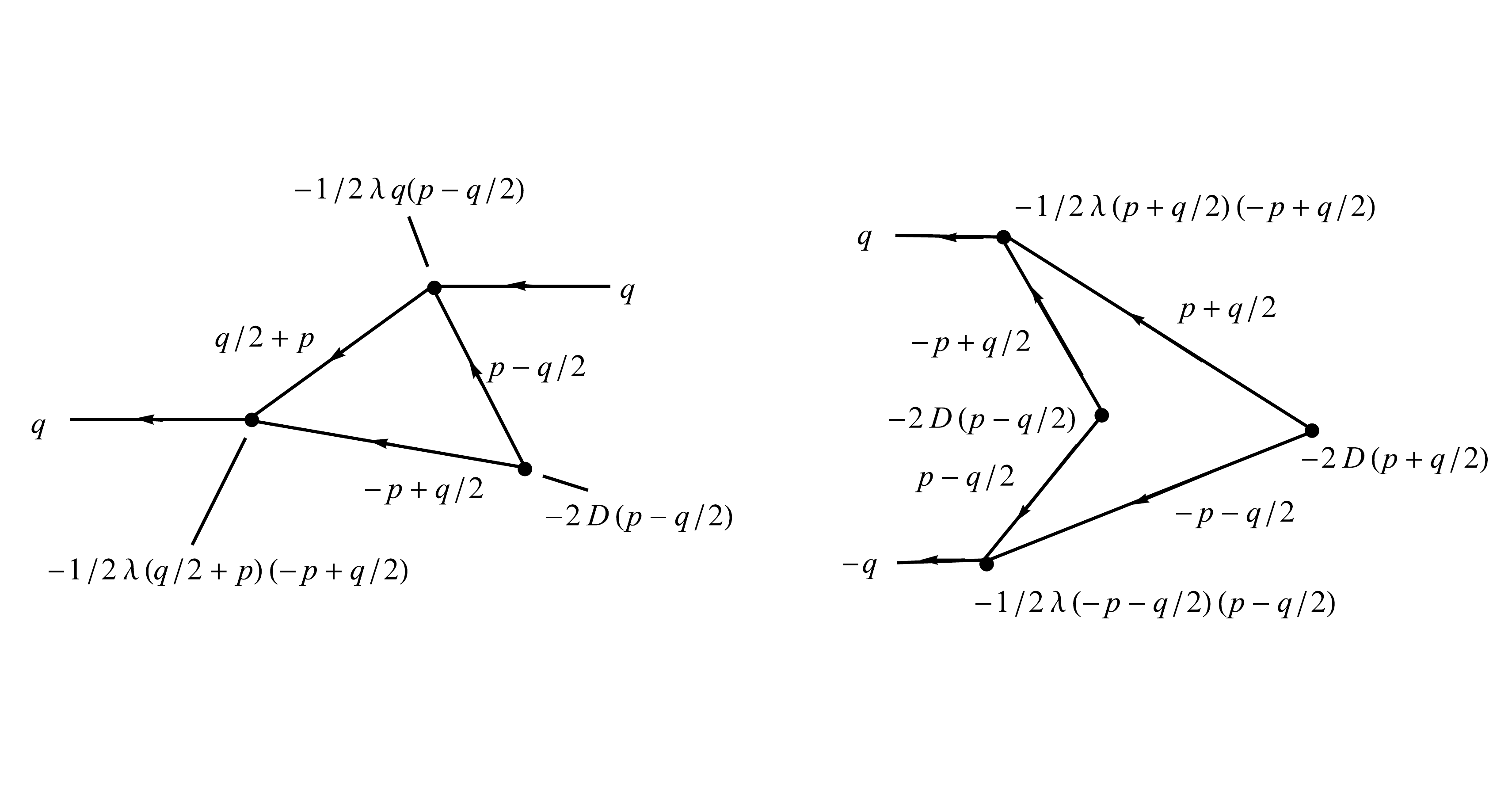}
 \caption{Feynman diagrams for the KPZ equation in a style following \cite{Taeuber2014_Book}. The left hand diagram shows the one-loop order expansion of the propagator vertex function $\Gamma_{\wt{h}h}$, whereas the right hand side depicts the one-loop order expansion of the noise vertex function $\Gamma_{\wt{h}\wt{h}}$. Here $q$ denotes the outer momentum and $p$ stands for the inner momentum. Note that a symmetrization has already been done, which leads to the noise amplitudes depending on $p\pm q/2$. For reasons of clarity the frequency components carried by each line corresponding to the different momenta are omitted.}
 \label{Feynman_Diagrams}
\end{figure*}
The vertex functions $\Gamma_{\wt{h}h}$ and $\Gamma_{\wt{h}\wt{h}}$ will be calculated to one-loop order in Fourier-space for an arbitrary noise with correlations of the form \eqref{General_Noise_Classification}.\par
Denoting the free propagator by $G_0(\kk)$ ($\kk=(q,\omega)$) and the self energy by $\Sigma(\kk)$, the analytic expressions for the diagrams in Fig. \ref{Feynman_Diagrams} are given by the Dyson equation $\Gamma_{\wt{h}h}=G_0(-\kk)^{-1}-\Sigma(\kk)$ \cite{Zinn_Justin2007,Amit1984,Mussardo2010} and the expansion of the noise vertex $\Gamma_{\wt{h}\wt{h}}=-2D(q)+\text{\textit{higher orders}}$. Using the usual Feynman rules (see e.g. \cite{Taeuber2014_Book}) and integrating out the inner frequencies one obtains\\\\
\begin{widetext}
\begin{align}
 \Gamma_{\wt{h}h}(\kk)&=i\omega+\nu\,q^2+\frac{\lambda^2}{2\,\nu^2}\int_p\frac{D(|p-q/2|^2)(q^2/2-q\cdot p)}{(q/2-p)^2}\frac{q^2/4-p^2}{\frac{i\omega}{2\nu}+q^2/4+p^2}+\mathcal{O}(\lambda^3),\label{Propagator_Vertex_Expansion_One_Loop}\\*
 \Gamma_{\wt{h}\wt{h}}(\kk)&=-2D(|q|^2)-\frac{\lambda^2}{2\,\nu^3}\int_p\,D(|p+q/2|^2)D(|p-q/2|^2)\Re\left[\frac{1}{\frac{i\omega}{2\nu}+q^2/4+p^2}\right]+\mathcal{O}(\lambda^3),\label{Noise_Vertex_Expansion_One_Loop}
\end{align}
\end{widetext}
where we used the following abbreviation: $\int_p\cdots=1/(2\pi)^d\int d^dp\cdots$.\par
Evaluating \eqref{Propagator_Vertex_Expansion_One_Loop} and \eqref{Noise_Vertex_Expansion_One_Loop} it is essential to avoid mixing ultraviolet and infrared divergences of the integrands. One way to keep those divergences separated is to introduce a so-called normalization point (NP). An indiscriminate, however very useful choice is given by \cite{Taeuber1994}
\begin{align}
 \frac{\omega}{2\,\nu}=\mu^2,\qquad q=0,
 \label{Normalization_Point}
\end{align}
where $\mu$ is an arbitrary momentum scale. One advantage of the choice in \eqref{Normalization_Point} is that evaluating the integrals in \eqref{Propagator_Vertex_Expansion_One_Loop} and \eqref{Noise_Vertex_Expansion_One_Loop} at $q=0$ yields the possibility of expanding the general noise amplitude $D(|p\pm q/2|^2)$ about $p$ for $|q|\ll1$. Hence to order $\mathcal{O}(|q|^2)$ the momentum-dependent noise amplitude reads
\begin{align}
 D\left(\left|p\pm\frac{q}{2}\right|^2\right)=D\left(|p|^2\right)\pm(p\cdot q)D^\prime\left(|p|^2\right)+\mathcal{O}(|q|^2).
 \label{Noise_Amplitude_Expansion}
\end{align}
Using the identities ($d$ is the spatial dimension) \cite{Taeuber1994,Taeuber2014_Book}
\begin{align*}
  \int_p\,(p\cdot q)^2h(|p|,|q|)&=\frac{q^2}{d}\int_p\,p^2h(|p|,|q|),\\
  \int_p\,p^2(p\cdot q)^2h(|p|,|q|)&=\frac{q^2}{d}\int_p\,p^4h(|p|,|q|)
\end{align*}
and inserting \eqref{Noise_Amplitude_Expansion} into \eqref{Propagator_Vertex_Expansion_One_Loop} implies at the NP 
\begin{align}
\begin{split}
  \left.\parderiv{\Gamma_{\wt{h}h}}{q^2}\right|_{q=0}&=\nu-\frac{\lambda^2}{4\,\nu^2}\frac{d-2}{d}\int_p\frac{D(|p|^2)}{i\mu^2+p^2}-\\
  &\qquad-\frac{\lambda^2}{2\,\nu^2}\frac{1}{d}\int_p\frac{p^2D^\prime(|p|^2)}{i\mu^2+p^2}.
  \end{split}
\label{Derivative_Propagator_Vertex_Expansion_One_Loop_NP}
\end{align}
The evaluation of \eqref{Noise_Vertex_Expansion_One_Loop} at the NP \eqref{Normalization_Point} leads with \eqref{Noise_Amplitude_Expansion} to
\begin{align}
 \Gamma_{\wt{h}\wt{h}}=-2D(|q|^2)-\frac{\lambda^2}{2\,\nu^3}\int_p\left[D(|p|^2)\right]^2\frac{p^2}{\mu^4+p^4}.
 \label{Noise_Vertex_Expansion_One_Loop_NP}
\end{align}
From \eqref{Derivative_Propagator_Vertex_Expansion_One_Loop_NP} and \eqref{Noise_Vertex_Expansion_One_Loop_NP} we obtain the renormalization factors 
\begin{align}
\begin{split}
 Z_\nu&=1-\frac{\lambda^2}{4\,\nu^3}\frac{d-2}{d}\int_p\frac{D(|p|^2)}{i\mu^2+p^2}-\\
 &\qquad\qquad-\frac{\lambda^2}{2\,\nu^3}\frac{1}{d}\int_p\frac{p^2D^\prime(|p|^2)}{i\mu^2+p^2},\end{split}\label{Z_Nu_General}\\
 Z_D&=1+\frac{1}{D(|q|^2)\Big|_{q=0}}\frac{\lambda^2}{4\,\nu^3}\int_p\left[D(|p|^2)\right]^2\frac{p^2}{\mu^4+p^4}.\label{Z_D_General}
\end{align}
These results allow us to compute the Wilson flow functions \cite{Janssen1999,Taeuber1994,Taeuber2014_Book}
\begin{align}
 \gamma_D&=\mu\parderiv{}{\mu}\ln Z_D,\label{Gamma_D_General}\\
 \gamma_\nu&=\mu\parderiv{}{\mu}\ln Z_\nu\label{Gamma_Nu_General},
\end{align}
where the derivative is taken while keeping $D$ and $\nu$ fixed. Likewise, the $\beta$-function is given by
\begin{align}
 \beta_g=\mu\parderiv{}{\mu}g_R,\label{Beta_g_General}
\end{align}
where 
\begin{align*}
 g_R=gZ_g\mu^{d-2}=gZ_DZ_\nu^{-3}\mu^{d-2}\sim\frac{\lambda^2\,D}{4\,\nu^3}\mu^{d-2}
\end{align*}
is a dimensionless effective coupling constant and $D=D(0)$ given in \eqref{Sinc_Noise_Correlation}.\par
The dimension of an effective coupling constant in the above form is (see e.g. \cite{Taeuber1994})
\begin{align}
 [g]=\left[\frac{\lambda^2\,D}{4\,\nu^3}\right]=\mu^{2-d}.
 \label{Effective_Coupling_Constant_Dimenison}
\end{align}
This explains why $g$ has to be multiplied by $\mu^{d-2}$ to render $g_R$ dimensionless.\par
With the flow functions \eqref{Gamma_D_General}--\eqref{Beta_g_General} a partial differential renormalization group equation can be formulated. This RG equation may be solved by using the method of characteristics, where a flow parameter $l$ and an $l$-dependent continuous momentum scale $\wt{\mu}(l)=\mu\,l$ is introduced. Those solutions are then used to formulate a KPZ-specific scaling relation for, say, the two point correlation function $C(q,\omega)$. This relation reads \cite{Taeuber1994}
\begin{align}
 C(\mu,D_R,\nu_R,g_R,q,\omega)=q^{-4-2\gamma_\nu^*+\gamma_D^*}\hat{C}\left(\frac{\omega}{q^{2+\gamma_\nu^*}}\right),
 \label{Explicit_Scaling_Correlation_Function}
\end{align}
where the superscript \lq$*$\rq{} indicates that the Wilson flow functions are evaluated at the stable IR fixed point. A detailed explanation of how the scaling form in \eqref{Explicit_Scaling_Correlation_Function} is obtained can be found e.g. in \cite{Taeuber1994,Taeuber2014_Book}. A comparison of \eqref{Explicit_Scaling_Correlation_Function} with the general scaling form for the KPZ two-point correlation function in Fourier-space (see e.g. \cite{Kardar_Parisi_Zhang1986,Taeuber1994,Family1985,Kosterlitz1989}), i.e.
\begin{align}
 C(q,\omega)=q^{-d-2\chi-z}\hat{C}\left(\frac{\omega}{q^z}\right),
 \label{General_Scaling_Correlation_Function}
\end{align}
leads to the following expressions for the dynamical exponent $z$ and the roughness exponent $\chi$ \cite{Taeuber1994,Taeuber2014_Book}:
\begin{align}
 z&=2+\gamma_\nu^*,\label{Dynamical_Exponent_General}\\
 \chi&=1-\frac{d}{2}+\frac{\gamma_\nu^*-\gamma_D^*}{2}.\label{Roughness_Exponent_General}
\end{align}
These general considerations will be used in the next part to obtain the critical exponents $z$ and $\chi$ for the explicit \lq sinc\rq{}-noise correlation.

\section{The KPZ Equation with \lq Sinc\rq{}-Noise Correlation}\label{The KPZ Equation with Sinc-Noise Correlation}

We now apply these results to the case of the \lq sinc\rq{}-type noise with the correlations
\begin{align}
\begin{split}
 \expval{\eta(q,\omega)\eta(q^\prime,\omega^\prime)}&=2D(|q|^2)\delta^d(q+q^\prime)\delta(\omega+\omega^\prime)\\
 &=2D\frac{\sin(\xi|q|)}{\xi|q|}\delta^d(q+q^\prime)\delta(\omega+\omega^\prime),
 \end{split}
 \label{Sinc_Noise_Correlation}
\end{align}
where $D$ is a constant noise amplitude, $q\in\mathds{R}^d$ and $\xi$ defines the scale of the \lq sinc\rq{}-profile.\par
For simplicity let us consider the case $d=1$. Here the noise distribution transformed back to real-space is a rectangle with size $2\xi\times D/\xi$ centered at $x=0$, which tends to $\delta(x)$ (white noise) in the limit $\xi\to0$ \cite{Kardar_Parisi_Zhang1986}.\par
The first step now is to calculate explicit expressions for the renormalization factors from \eqref{Z_Nu_General} and \eqref{Z_D_General} for the driving force \eqref{Sinc_Noise_Correlation}. Thus with 
\begin{align*}
 D(|p|^2)&=D\frac{\sin(\xi|p|)}{\xi|p|}\\
 \intertext{and}
 D^\prime(|p|^2)&=\deriv{D}{(|p|^2)}
\end{align*}
the renormalization factors read in $d=1$ to one-loop order
\begin{align}
\begin{split}
  Z_\nu&=1+\frac{D\,\lambda^2}{4\,\nu^3}\frac{1}{\pi}\left[2\int_0^\infty dp\,\frac{\sin(\xi p)}{\xi p(i\mu^2+p^2)}-\right.\\
  &\left.\qquad-\int_0^\infty dp\,\frac{\cos(\xi p)}{i\mu^2+p^2}\right],\end{split}\label{Z_Nu_Sinc}\\*
  Z_D&=1+\frac{D\,\lambda^2}{4\,\nu^3}\frac{1}{\pi}\int_0^\infty dp\,\frac{\sin^2(\xi p)}{\xi^2(\mu^4+p^4)}.\label{Z_D_Sinc}
\end{align}
The integrals occurring in \eqref{Z_Nu_Sinc} and \eqref{Z_D_Sinc} can be computed by employing the Residue theorem, which leads to
\begin{align}
\begin{split}
 &\qquad Z_\nu=1+\frac{D\,\lambda^2}{4\,\nu^3}\,\frac{e^{-\frac{1}{\sqrt{2}}\xi\,\mu}}{\xi\,\mu^2}\times\\
 &\times\left[\sin\left(\frac{1}{\sqrt{2}}\xi\,\mu\right)\left(1+\frac{\xi\,\mu}{2\sqrt{2}}\right)-\frac{\xi\,\mu}{2\sqrt{2}}\cos\left(\frac{1}{\sqrt{2}}\xi\,\mu\right)\right],\end{split}\label{Z_Nu_Sinc_Evaluated}\\
 \begin{split}
 &\qquad Z_D=1+\frac{D\,\lambda^2}{4\,\nu^3}\,\frac{e^{-\sqrt{2}\xi\,\mu}}{4\sqrt{2}\xi^2\,\mu^3}\times\\
 &\times\left[e^{\sqrt{2}\xi\,\mu}-\left(\sin(\sqrt{2}\xi\,\mu)+\cos(\sqrt{2}\xi\,\mu)\right)\right].\end{split}\label{Z_D_Sinc_Evaluated}
\end{align}
In Appendix \ref{Explicit Evaluation of the Renormalization Factors} the derivation of these formulas is explained in greater detail.

\subsection{Small Correlation Length Expansion}\label{Small Xi Expansion}

Let us now focus on small correlation lengths $\xi\ll1$ and expand \eqref{Z_Nu_Sinc_Evaluated} and \eqref{Z_D_Sinc_Evaluated} in $\xi$ up to order $\mathcal{O}(\xi^2)$.\\
Introducing the effective coupling constants \cite{Medina1989}
\begin{alignat}{2}
 g&=\frac{D\,\lambda^2}{4\,\nu^3},\qquad\quad\:\: g_R&&=\frac{g\,Z_g}{2\sqrt{2}\mu};\label{Effective_Coupling_Constant_g}\\
 g_\xi&=\frac{D\,\xi^2\,\lambda^2}{4\,\nu^3},\qquad g_{\xi,R}&&=\frac{g_\xi\,Z_g\,\mu}{2\sqrt{2}}.\label{Effective_Coupling_Constant_g_Xi}
\end{alignat}
with $Z_g=Z_DZ_\nu^{-3}$ the one-loop integrals simplify:
\begin{align}
 Z_\nu&=1+\frac{g}{2\sqrt{2}\mu}-\frac{g_\xi\,\mu}{12\sqrt{2}},\label{Z_Nu_Sinc_Expansion}\\
 Z_D&=1+\frac{g}{2\sqrt{2}\mu}-\frac{D\,\lambda^2}{12\,\nu^3}\xi+\frac{g_\xi\,\mu}{6\sqrt{2}}.\label{Z_D_Sinc_Expansion}
\end{align}
With the renormalized dimensionless effective coupling constants from \eqref{Effective_Coupling_Constant_g}, \eqref{Effective_Coupling_Constant_g_Xi} and using \eqref{Beta_g_General} one obtains the flow equations
\begin{align}
 \beta_g&=g_R\left[2g_R+\frac{5}{6}g_{\xi,R}-1\right],\label{Beta_g_Sinc_Expansion}\\
 \beta_{g_\xi}&=g_{\xi,R}\left[2g_R+\frac{5}{6}g_{\xi,R}+1\right].\label{Beta_g_Xi_Sinc_Expansion}
\end{align}
Solving \eqref{Beta_g_Sinc_Expansion} and \eqref{Beta_g_Xi_Sinc_Expansion} for their fixed points $(g^*,g_\xi^*)$ yields three different possible solutions, namely
\begin{align}
 \begin{split}
  (g_R^*,g_{\xi,R}^*)=\begin{cases} (0,0)\quad\text{Gaussian},\\(0,-\frac{6}{5}),\\(\frac{1}{2},0)\quad\text{KPZ}. \end{cases}
 \end{split}
\label{Fixed_Points_Sinc_Expansion}
\end{align}
The second one is nonphysical, since $g_{\xi,R}<0$, and thus there are two valid fixed points for the KPZ equation stirred by \lq sinc\rq{}-type noise with $\xi\ll1$.\\
To determine the stability of the two fixed points we carry out a linear stability analysis via the Jacobian of the two flow functions \eqref{Beta_g_Sinc_Expansion} and \eqref{Beta_g_Xi_Sinc_Expansion}, i.e.
\begin{align}
\begin{split}
 \mathcal{J}&=\begin{pmatrix} \partial_{g_R}\beta_g & \partial_{g_{\xi,R}}\beta_{g} \\ \partial_{g_R}\beta_{g_\xi} & \partial_{g_{\xi,R}}\beta_{g_\xi}\end{pmatrix}\\
 &=\begin{pmatrix} 4g_R+5/6\,g_{\xi,R}-1 & 5/6\,g_R \\ 2g_{\xi,R} & 2g_R+5/3\,g_{\xi,R}+1 \end{pmatrix}.
 \end{split}
 \label{Jacobian_Sinc_Expansion}
\end{align}
By evaluating \eqref{Jacobian_Sinc_Expansion} at the respective fixed points it turns out that for the Gaussian fixed point $\mathcal{J}$ is indefinite and for the KPZ fixed point $\mathcal{J}$ is positive definite. Since the condition for asymptotic stability in this framework is positive definiteness of \eqref{Jacobian_Sinc_Expansion}, only the KPZ fixed point is stable in the infrared limit and the Gaussian fixed point is unstable.\par
\begin{figure}[t]
 \centering
 \includegraphics[width=0.9\linewidth]{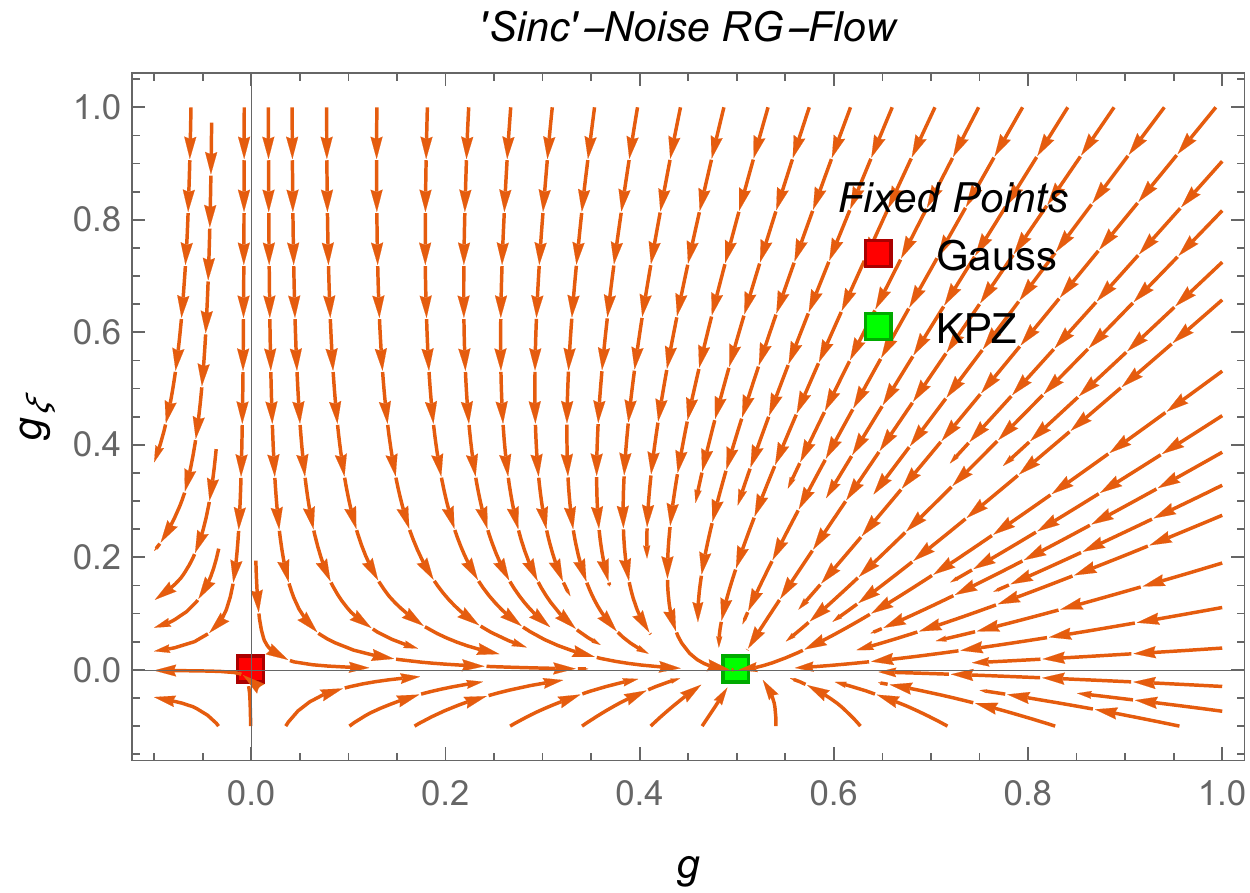}
 \caption{Renormalization group flow of the two effective coupling constants \eqref{Effective_Coupling_Constant_g} and \eqref{Effective_Coupling_Constant_g_Xi} for small values of the noise correlation length $\xi$ in $d=1$ spatial dimension. As can be seen the only stable infrared fixed point is the KPZ fixed point at $(g,g_\xi)=(1/2,0)$.}
 \label{Sinc_Flow}
\end{figure}
To provide a simple graphical representation of the occurring renormalization group flow Fig. \ref{Sinc_Flow} was plotted in Wilson's picture \cite{Wilson1975}. Parametrizing the scale transformation from the field-theoretic to Wilson's representation by (see e.g. \cite{Taeuber1994})
\begin{align}
 l_W=-\ln l
\end{align}
one obtains the following flow equations
\begin{align}
 \deriv{\wt{g}(l_W)}{l_W}&=-\beta_g\stackrel{\eqref{Beta_g_Sinc_Expansion}}{=}-\wt{g}(l_W)\left[2\wt{g}(l_W)+\frac{5}{6}\wt{g}_\xi(l_W)-1\right],\label{Beta_g_Sinc_Expansion_Wilson}\\
 \deriv{\wt{g}_\xi(l_W)}{l_W}&=-\beta_\xi\stackrel{\eqref{Beta_g_Xi_Sinc_Expansion}}{=}-\wt{g}_\xi(l_W)\left[2\wt{g}(l_W)+\frac{5}{6}\wt{g}_\xi(l_W)+1\right].\label{Beta_g_Xi_Sinc_Expansion_Wilson}
\end{align}
The corresponding RG flow is displayed in Fig. \ref{Sinc_Flow}.\par
The critical exponents $z$ and $\chi$ are obtained via \eqref{Dynamical_Exponent_General} and \eqref{Roughness_Exponent_General}. Here the fixed point values of the Wilson flow functions,
\begin{align}
 \gamma_\nu&=-g_R-\frac{g_{\xi,R}}{6}+\mathcal{O}(g_R^2,g_{\xi,R}^2,g_Rg_{\xi,R}),\label{Gamma_Nu_Explicit_Sinc_Expansion}\\
 \gamma_D&=-g_R+\frac{g_{\xi,R}}{3}+\mathcal{O}(g_R^2,g_{\xi,R}^2,g_Rg_{\xi,R}),\label{Gamma_D_Explicit_Sinc_Expansion},
\end{align}
are given by
\begin{align}
 \gamma_\nu^*&=\gamma_\nu(g_R=1/2,g_{\xi,R}=0)=-\frac{1}{2},\label{Gamma_Nu_Explicit_Sinc_Expansion_Fixed_Point}\\
 \gamma_D^*&=\gamma_D(g_R=1/2,g_{\xi,R}=0)=-\frac{1}{2}.\label{Gamma_D_Explicit_Sinc_Expansion_Fixed_Point}
\end{align}
Hence the dynamical exponent $z$ and the roughness exponent $\chi$ read
\begin{align}
 z=\frac{3}{2},\qquad\chi=\frac{1}{2}.\label{Critical_Exponents_Sinc_Expansion}
\end{align}
They are the same as those in the white-noise case and confirm the KPZ exponent identity $z+\chi=2$ (see e.g. \cite{Kardar_Parisi_Zhang1986,Medina1989,Janssen1999,Taeuber1994,HalpinHealy1995}).

\subsection{Arbitrary Correlation Length Calculation}\label{Arbitrary Xi Calculation}

In the following we show that the same result can be derived for arbitrary correlation lengths $\xi$ in $d=1$ dimensions, although the calculations are technically more involved.\par
Inserting \eqref{Z_Nu_Sinc_Evaluated}, \eqref{Z_D_Sinc_Evaluated} into \eqref{Gamma_Nu_General}, \eqref{Gamma_D_General} and expanding to lowest order in the effective coupling constant $g=D\lambda^2/(4\nu^3)$, the Wilson flow functions $\gamma_i$ can be written down as
\begin{widetext}
\begin{align}
 \begin{split}
  \gamma_\nu&=\mu\parderiv{\ln Z_\nu}{\mu}=-\frac{g_R}{Z_\nu}\frac{e^{-\frac{1}{\sqrt{2}}\xi\,\mu}}{\xi\,\mu}\left[\left(3\xi\,\mu+4\sqrt{2}\right)\sin\frac{\xi\,\mu}{\sqrt{2}}-\xi\,\mu\left(\sqrt{2}\xi\,\mu+3\right)\cos\frac{\xi\,\mu}{\sqrt{2}}\right]\\
  &=-g_R\frac{e^{-\frac{1}{\sqrt{2}}\xi\,\mu}}{\xi\,\mu}\left[\left(3\xi\,\mu+4\sqrt{2}\right)\sin\frac{\xi\,\mu}{\sqrt{2}}-\xi\,\mu\left(\sqrt{2}\xi\,\mu+3\right)\cos\frac{\xi\,\mu}{\sqrt{2}}\right]+\mathcal{O}(g_R^2),  \label{Gamma_Nu_Explicit}
 \end{split}\\
 \begin{split}
  \gamma_D&=\mu\parderiv{\ln Z_D}{\mu}=\frac{g_R}{Z_D}\frac{e^{-\sqrt{2}\xi\,\mu}}{2\xi^2\,\mu^2}\left[\left(2\sqrt{2}\xi\,\mu+3\right)\sin\sqrt{2}\xi\,\mu-3\left(e^{\sqrt{2}\xi\,\mu}-\cos\sqrt{2}\xi\,\mu\right)\right]\\
  &=g_R\frac{e^{-\sqrt{2}\xi\,\mu}}{2\xi^2\,\mu^2}\left[\left(2\sqrt{2}\xi\,\mu+3\right)\sin\sqrt{2}\xi\,\mu-3\left(e^{\sqrt{2}\xi\,\mu}-\cos\sqrt{2}\xi\,\mu\right)\right]+\mathcal{O}(g_R^2),
  \label{Gamma_D_Explicit}
 \end{split}
\end{align}
\end{widetext}
where we introduced the dimensionless form of the renormalized couplings
\begin{align}
 g_R=\frac{g\,Z_g}{2\sqrt{2}\mu},\qquad Z_g=Z_DZ_\nu^{-3}.
 \label{Effective_Coupling_Constant_Renormalized}
\end{align}
The corresponding $\beta$-function \eqref{Beta_g_General} reads 
\begin{align}
 \beta_g=g_R\left[g_R\left(\wt{\gamma}_D-3\wt{\gamma}_\nu\right)-1\right],
 \label{Beta_G_Explicit}
\end{align}
where $\wt{\gamma_i}=\gamma_i/g_R$ and $\gamma_\nu$, $\gamma_D$ are taken from \eqref{Gamma_Nu_Explicit}, \eqref{Gamma_D_Explicit}, respectively.\par
Again the flow of the effective coupling constant is modeled via the flow parameter $l$ used for the solution of the RG equations by the method of characteristics. This leads to a continuous momentum scale $\wt{\mu}(l)$, effective coupling constant $\wt{g}(l)$ and thus to an $l$-dependent flow equation (see e.g. \cite{Taeuber1994,Janssen1999})
\begin{align}
 \beta_g(l)=l\deriv{\wt{g}(l)}{l}.
 \label{Continuous_Beta_Function}
\end{align}
Hence a fixed point is characterized by $\beta_g(l)=0$. Applying this fixed point condition to \eqref{Beta_G_Explicit} and solving for $g_R$ leads to two separate infrared fixed point solutions $g_{R,i}^*$:
\begin{align}
 g_{R,1}^*&=0,\label{Trivial_Fixed_Point}\\*
 g_{R,2}^*&=\lim_{l\to0}\frac{1}{\wt{\gamma}_D(l)-3\wt{\gamma}_\nu(l)}\label{Non_Trivial_Fixed_Point_General}.
\end{align}
Here \eqref{Trivial_Fixed_Point} represents the trivial Gaussian fixed point while the second solution in the limit $l\to0$ \cite{Taeuber2014_Book} yields the nontrivial KPZ fixed point,
\begin{align}
 g_{R,2}^*=\lim_{l\to0}\frac{1}{\wt{\gamma}_D(l)-3\wt{\gamma}_\nu(l)}=\frac{1}{2}.
 \label{KPZ_Fixed_Point}
\end{align}
Again the fixed points are stable, if $d\beta_g/dg_R>0$. Since \eqref{Gamma_Nu_Explicit}, \eqref{Gamma_D_Explicit}, \eqref{Beta_G_Explicit} imply that
\begin{align}
 \beta_g^\prime=\deriv{\beta_g(l)}{\wt{g}(l)}=2\wt{g}(l)\left(\wt{\gamma}_D(l)-3\wt{\gamma}_\nu(l)\right)-1\stackrel{l\to0}{=}4g_R-1,
 \label{Fixed_Point_Stability_Condition}
\end{align}
we find that:
\begin{alignat*}{4}
 g_R^*&=0 &&:\qquad\beta_g^\prime &&=-1<0\quad&&\Longrightarrow\text{unstable},\\
 g_R^*&=\frac{1}{2} &&:\qquad\beta_g^\prime &&=1>0\quad&&\Longrightarrow\text{stable}.
\end{alignat*}
Hence there is one stable infrared fixed point, $g_R^*=1/2$, at which the critical exponents of the KPZ universality class can be calculated.\par
We obtain the critical exponents in $d=1$ dimensions again as
\begin{align}
 z&=2+\gamma_\nu^*=2-\frac{1}{2}=\frac{3}{2},\label{Dynamical_Exponent_Explicit}\\
 \chi&=\frac{1}{2}+\frac{-\frac{1}{2}+\frac{1}{2}}{2}=\frac{1}{2}.\label{Roughness_Exponent_Explicit}
\end{align}

\section{Discussion}\label{Discussion}

In the present work we have studied the field-theoretic DRG of the KPZ equation for correlated noise of \lq sinc\rq{}-type which is characterized by a finite correlation ~~~~length $\xi$.\par
The fixed points of the KPZ-DRG flow have been calculated in two different manners, namely first for small correlation lengths $\xi$ and via two effective coupling constants, $g$ and $g_\xi$ (see \autoref{Small Xi Expansion} and \eqref{Effective_Coupling_Constant_g}, \eqref{Effective_Coupling_Constant_g_Xi}) and then, using only one effective coupling constant $g$ (see \autoref{Arbitrary Xi Calculation}), for arbitrary values of $\xi$. Both methods yield the same results, i.e., an unstable Gaussian fixed point (see \eqref{Fixed_Points_Sinc_Expansion}, \eqref{Trivial_Fixed_Point}) and the stable KPZ fixed point (see \eqref{Fixed_Points_Sinc_Expansion}, \eqref{KPZ_Fixed_Point}).\par
It might be argued, that the second method is somewhat redundant since the \lq small\rq{}-$\xi$ expansion can also be interpreted to be valid for arbitrary values of $\xi$ in the infrared limit as in this regime $\mu\to0$ and hence $\xi\,\mu\eqqcolon\epsilon\ll1$. The expansion would then be done for the parameter $\epsilon$. Nevertheless, the method used in \autoref{Arbitrary Xi Calculation} is a reassuring confirmation of the results obtained in \autoref{Small Xi Expansion}.\par
Building on these fixed points, the critical exponents characterizing the KPZ universality class, i.e. the dynamical exponent $z=3/2$ and the roughness exponent $\chi=1/2$, were calculated (see \eqref{Critical_Exponents_Sinc_Expansion} and \eqref{Dynamical_Exponent_Explicit}, \eqref{Roughness_Exponent_Explicit}). The values obtained for $z$ and $\chi$ coincide with the \lq standard\rq{} KPZ exponents in one spatial dimension, where the system is excited by purely white noise (see e.g.\cite{Kardar_Parisi_Zhang1986,Taeuber1994}).\par
Hence, for every finite noise correlation length $\xi$ the system behaves to one-loop order as if it was stirred by the \lq standard\rq{} uncorrelated Gaussian noise from \eqref{White_Noise}.\par
This result corresponds nicely with the numerical findings of \cite{Canet2017}, where a different spatial noise correlation was analyzed. The authors there found that for small values of the noise correlation length the KPZ equation acted like it was driven by pure white noise.\par
Combining the findings of \cite{Canet2017} with the present ones, found for different noise functions and by different methods, we arrive at the conjecture that the large-scale KPZ dynamics is independent of the details of the noise structure, provided that the correlation length $\xi$ is finite.

\appendix
\section{Explicit Evaluation of the Renormalization Factors}\label{Explicit Evaluation of the Renormalization Factors}

To obtain \eqref{Z_Nu_Sinc_Evaluated}, \eqref{Z_D_Sinc_Evaluated} from the expressions in \eqref{Z_Nu_Sinc}, \eqref{Z_D_Sinc}, respectively, we use the Residue theorem. To this end the integrals are first rewritten in a more easily accessible form.

\subsection{Evaluation of Eq. \eqref{Z_Nu_Sinc}}\label{Evaluation Of Z_Nu_Sinc}
\begin{figure}[t]
 \centering
 \includegraphics[width=0.9\linewidth]{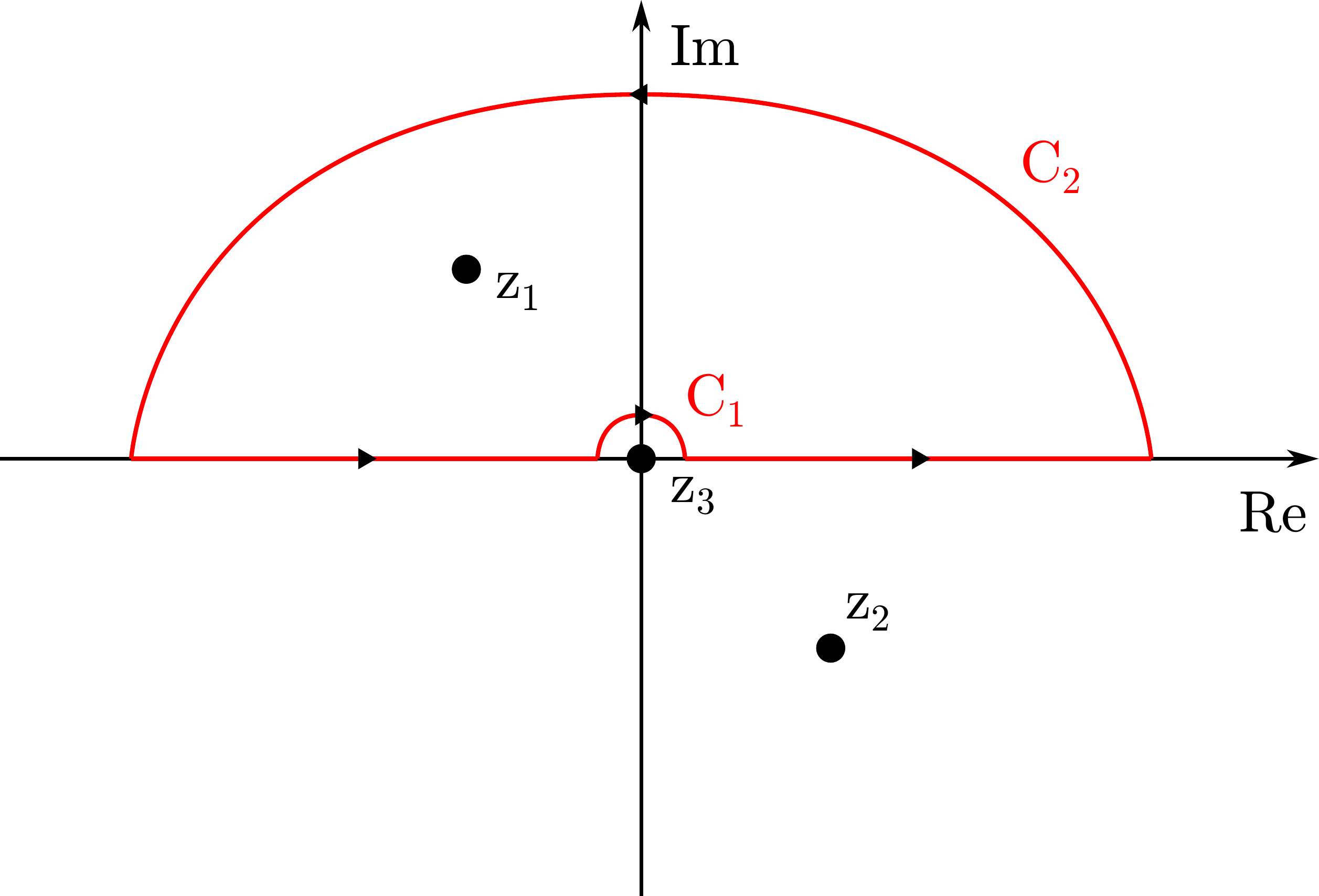}
 \caption{Integration contour $C$ for the evaluation of \eqref{Original_Int_1}. $C_1$ and $C_2$ are circles about $z=0$ with radius $\epsilon$ and $R$, respectively.}
 \label{Res_1}
\end{figure}

The first integral needed for the calculation of $Z_\nu$ reads
\begin{align}
 \int_0^\infty dp\,\frac{\sin(\xi\,p)}{\xi\,p(i\mu^2+p^2)}.
 \label{Original_Int_1}
\end{align}
This may be rewritten as
\begin{align}
 \int_0^\infty dp\,\frac{\sin(\xi\,p)}{\xi\,p(i\mu^2+p^2)}=-\frac{i}{2}\int_{-\infty}^\infty dp\,\frac{e^{i\xi\,p}}{\xi\,p(i\mu^2+p^2)}
 \label{Original_Int_1_Rewritten}
\end{align}
The integrand on the r.h.s. in \eqref{Original_Int_1_Rewritten} has three simple poles which are given by
$z_{1/2}=\pm\mu\, e^{i3\pi/4}$ and $z_3=0$. Those and the chosen integration contour are shown in Fig. \ref{Res_1}.
Hence the residue theorem yields
\begin{align*}
 &\quad\int_C dz\,\frac{e^{i\xi z}}{\xi z(i\mu^2+z^2)}\\*
 &=\int_{-R}^{-\epsilon}dz\,\frac{e^{i\xi z}}{\xi z\left(i\mu^2+z^2\right)}+\int_{C_1}dz\,\frac{e^{i\xi z}}{\xi z\left(i\mu^2+z^2\right)}+\\*
 &\qquad+\int_\epsilon^R dz\,\frac{e^{i\xi z}}{\xi z\left(i\mu^2+z^2\right)}+\int_{C_2}dz\,\frac{e^{i\xi z}}{\xi z\left(i\mu^2+z^2\right)}\\
 &=2\,\pi\,i\lim_{z\to\mu\,e^{i3\pi/4}}\frac{\left(z-\mu\,e^{i3\pi/4}\right)e^{i\xi z}}{\xi z\left(z-\mu\,e^{i3\pi/4}\right)\left(z+\mu\,e^{i3\pi/4}\right)}\\
 &=-\frac{\pi\,e^{-\frac{1}{\sqrt{2}}\xi\mu}}{\xi\mu^2}\left[\cos\left(\frac{1}{\sqrt{2}}\xi\mu\right)-i\sin\left(\frac{1}{\sqrt{2}}\xi\mu\right)\right].
\end{align*}
To obtain the integral on the real axis from minus to plus infinity, the contributions of the integrals over the two circular paths have to be computed. Therefore the parametrization 
\begin{align*}
 z=\epsilon\,e^{i\varphi}\quad\Leftrightarrow\quad dz=i\,\epsilon\,e^{i\varphi}d\varphi
\end{align*}
is used, which yields for the integral over $C_1$ with $\epsilon\to0$:
\begin{align*}
 &\lim_{\epsilon\to0}\int_{C_1}dz\,\frac{e^{i\xi z}}{\xi z\left(i\mu^2+z^2\right)}\\
 &\quad=-\lim_{\epsilon\to0}\int_0^\pi d\varphi\,\frac{i\,e^{i\xi \epsilon e^{i\varphi}}}{\xi \left(i\mu^2+\epsilon^2\,e^{2i\varphi}\right)}\\
 &\quad=-\int_0^\pi d\varphi\,\lim_{\epsilon\to0}\frac{i\,e^{i\xi \epsilon e^{i\varphi}}}{\xi \left(i\mu^2+\epsilon^2\,e^{2i\varphi}\right)}=-\frac{\pi}{\xi\,\mu^2}
\end{align*}
For the integration over the contour $C_2$ a similar parametrization is used
\begin{align}
 z=R\,e^{i\varphi}\quad\Leftrightarrow\quad dz=i\,R\,e^{i\varphi}d\varphi.
 \label{Parametrization}
\end{align}
The contribution from this integral vanishes for $R\to\infty$:
\begin{align*}
 &\lim_{R\to\infty}\left|\int_{C_2}dz\,\frac{e^{i\xi z}}{\xi z\left(i\mu^2+z^2\right)}\right|\\
 &\quad=\lim_{R\to\infty}\left|\int_0^\pi d\varphi\,\frac{i\,R\,e^{i\xi Re^{i\varphi}}e^{i\varphi}}{\xi Re^{i\varphi}\left(i\mu^2+R^2e^{2i\varphi}\right)}\right|\\
 &\quad\leq\lim_{R\to\infty}\int_0^\pi d\varphi\,\frac{\left|e^{i\xi Re^{i\varphi}}\right|}{\left|\xi \left(i\mu^2+R^2e^{2i\varphi}\right)\right|}\\
 &\quad=\lim_{R\to\infty}\int_0^\pi d\varphi\,\frac{e^{-\xi R\sin\varphi}}{\xi R^2\left|i\frac{\mu^2}{R^2}+e^{2i\varphi}\right|}=0,
\end{align*}
since $\sin\varphi>0$ for $0<\varphi<\pi$. Thus in the limits $R\to\infty$ and $\epsilon\to0$ the residue theorem results in
\begin{align*}
 &\int_{-\infty}^\infty dz\,\frac{e^{i\xi z}}{\xi z\left(i\mu^2+z^2\right)}=\frac{\pi\,e^{-\frac{1}{\sqrt{2}}\xi\mu}}{\xi\,\mu^2}\left[e^{\frac{1}{\sqrt{2}}\xi\mu}-\right.\\
 &\left.\qquad\qquad\qquad-\left(\cos\left(\frac{1}{\sqrt{2}}\xi\mu\right)-i\sin\left(\frac{1}{\sqrt{2}}\xi\mu\right)\right)\right].
\end{align*}
The integral from \eqref{Original_Int_1} is therefore given by
\begin{align}
\begin{split}
  &\int_0^\infty dp\,\frac{\sin(\xi p)}{\xi p\left(i\mu^2+p^2\right)}=\pi\frac{e^{-\frac{1}{\sqrt{2}}\xi\mu}}{2\,\xi\,\mu^2}\times\\
  &\quad\times\left[\sin\left(\frac{1}{\sqrt{2}}\xi\mu\right)+i\left(\cos\left(\frac{1}{\sqrt{2}}\xi\mu\right)-e^{\frac{1}{\sqrt{2}}\xi\mu}\right)\right].
  \end{split}
  \label{Original_Int_1_Final}
\end{align}
\par
\begin{figure}[t]
 \centering
 \includegraphics[width=0.9\linewidth]{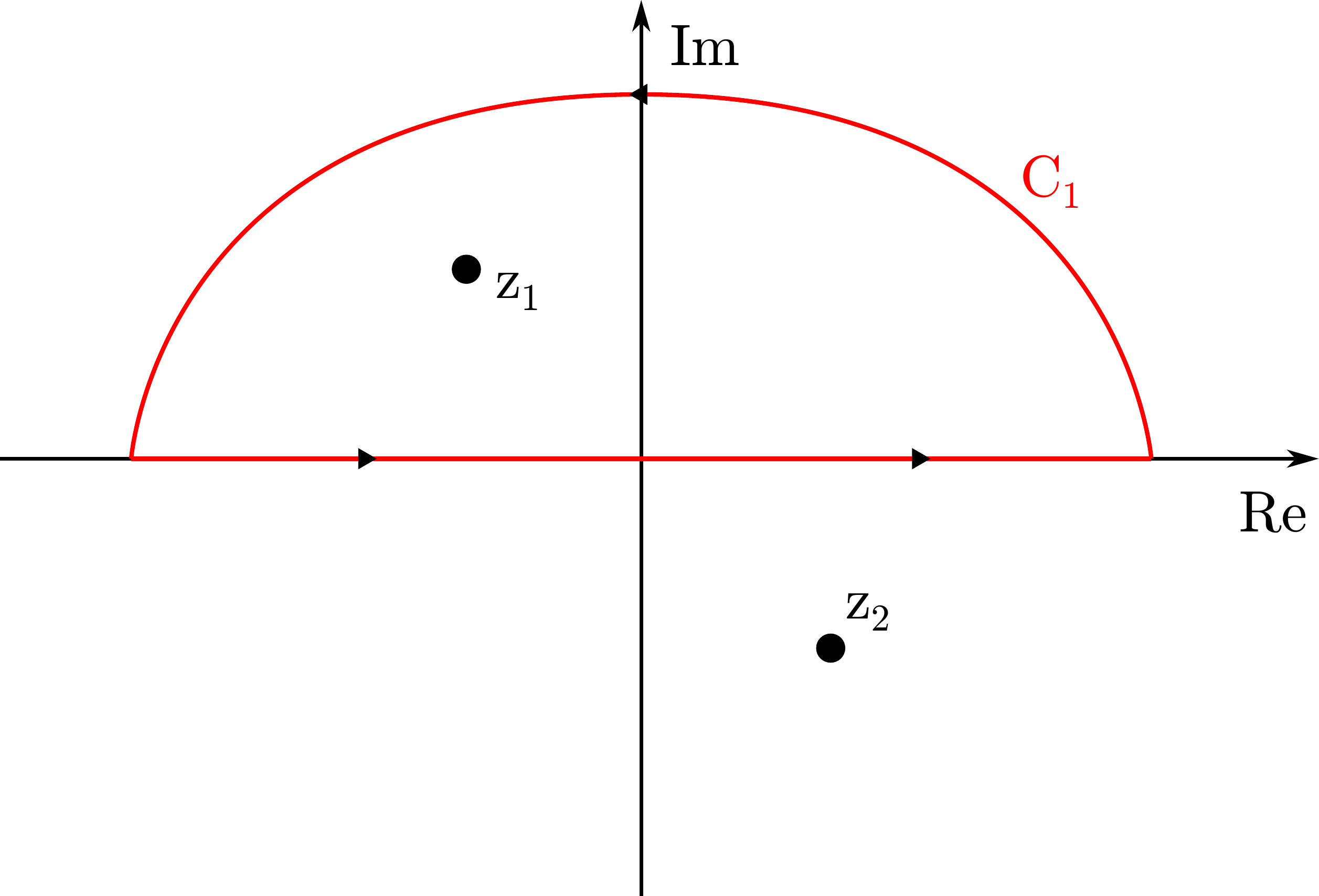}
 \caption{This diagram shows the contour $C$ of integration for \eqref{Original_Int_2_Rewritten}. $C_1$ denotes a circle about $z=0$ with radius $R$.}
 \label{Res_2}
\end{figure}

The second integral needed for the evaluation of \eqref{Z_Nu_Sinc} is given by
\begin{align}
 \int_0^\infty dp\,\frac{\cos(\xi\,p)}{i\mu^2+p^2}.
 \label{Original_Int_2}
\end{align}\\
As for the calculation of \eqref{Original_Int_1}, the integral will be rewritten according to
\begin{align}
 \int_0^\infty dp\,\frac{\cos(\xi\,p)}{i\mu^2+p^2}=\frac{1}{2}\int_{-\infty}^\infty dp\,\frac{e^{i\xi\,p}}{i\mu^2+p^2}.
 \label{Original_Int_2_Rewritten}
\end{align}
The integrand on the r.h.s. of \eqref{Original_Int_2_Rewritten} has two simple poles at $z_{1/2}=\pm\,\mu\,e^{i3\pi/4}$. For the integration contour shown in Fig. \ref{Res_2}, the residue theorem leads to
\begin{align*}
 &\int_Cdz\,\frac{e^{i\xi z}}{i\mu^2+z^2}=\int_{-R}^Rdz\,\frac{e^{i\xi z}}{i\mu^2+z^2}+\int_{C_1}dz\,\frac{e^{i\xi z}}{i\mu^2+z^2}\\
 &\quad=2\,\pi\,i\lim_{z\to\mu e^{i3\pi/4}}\frac{\left(z-\mu\,e^{i3\pi/4}\right)e^{i\xi z}}{\left(z-\mu\,e^{i3\pi/4}\right)\left(z+\mu\,e^{i3\pi/4}\right)}\\
 &\quad=\frac{\pi\,e^{-\frac{1}{\sqrt{2}}\xi\mu}}{\sqrt{2}\,\mu}\left(\cos\left(\frac{1}{\sqrt{2}}\xi\mu\right)-\sin\left(\frac{1}{\sqrt{2}}\xi\mu\right)-\right.\\
 &\left.\qquad-i\left[\cos\left(\frac{1}{\sqrt{2}}\xi\mu\right)+\sin\left(\frac{1}{\sqrt{2}}\xi\mu\right)\right]\right)
\end{align*}
Choosing again the parametrization \eqref{Parametrization} it is readily shown that its contribution vanishes for $R\to\infty$:
\begin{align*}
 &\qquad\lim_{R\to\infty}\left|\int_{C_1}dz\,\frac{e^{i\xi z}}{i\mu^2+z^2}\right|\\
 &=\lim_{R\to\infty}\left|\int_0^\pi d\varphi\,\frac{i\,R\,e^{i\xi Re^{i\varphi}}e^{i\varphi}}{i\mu^2+R^2\,e^{2i\varphi}}\right|\\
 &\leq\lim_{R\to\infty}\int_0^\pi d\varphi\,\frac{\left|i\,R\,e^{i\xi Re^{i\varphi}}e^{i\varphi}\right|}{\left|i\mu^2+R^2\,e^{2i\varphi}\right|}\\
 &=\lim_{R\to\infty}\int_0^\pi d\varphi\,\frac{e^{-\xi R\sin\varphi}}{R\left|\frac{i\mu^2}{R^2}+e^{2i\varphi}\right|}=0
\end{align*}
Hence the sought integral reads
\begin{align}
\begin{split}
  &\int_0^\infty dp\,\frac{\cos(\xi p)}{i\mu^2+p^2}\\
  &\quad=\pi\frac{e^{-\frac{1}{\sqrt{2}}\xi\mu}}{2\,\sqrt{2}\,\mu}\left(\cos\left(\frac{1}{\sqrt{2}}\xi\mu\right)-\sin\left(\frac{1}{\sqrt{2}}\xi\mu\right)-\right.\\
  &\left.\qquad-i\left[\cos\left(\frac{1}{\sqrt{2}}\xi\mu\right)+\sin\left(\frac{1}{\sqrt{2}}\xi\mu\right)\right]\right).
  \end{split}
  \label{Original_Int_2_Final}
\end{align}
\par
Taking the real parts \cite{Janssen1999} of \eqref{Original_Int_1_Final} and \eqref{Original_Int_2_Final} and inserting the results into \eqref{Z_Nu_Sinc} leads to the expression in \eqref{Z_Nu_Sinc_Evaluated}.

\subsection{Evaluation of Eq. \eqref{Z_D_Sinc}}\label{Evaluation Of Z_D_Sinc}

\begin{figure}[t]
 \centering
 \includegraphics[width=0.9\linewidth]{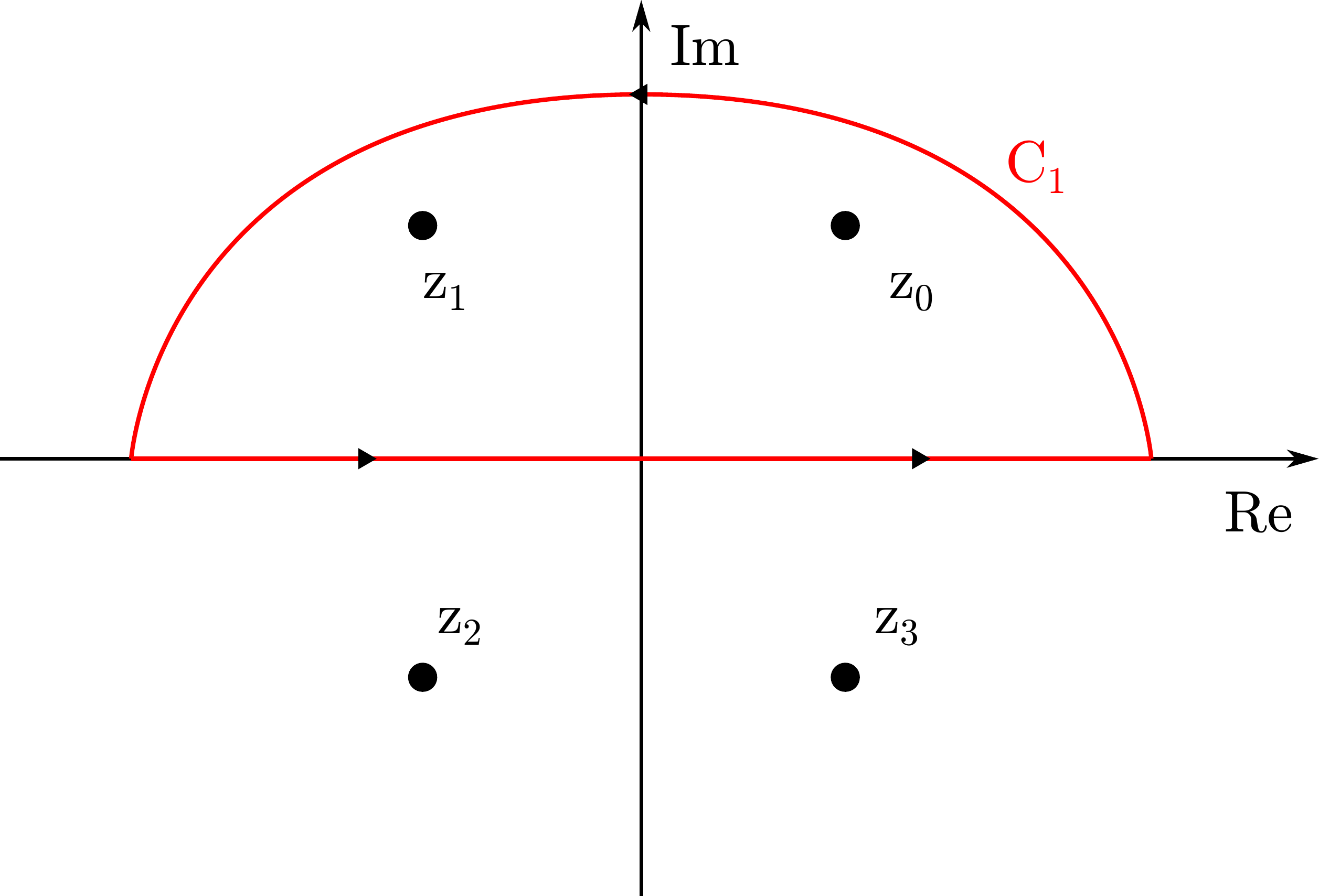}
 \caption{Representation of the zeros and the contour $C$ of integration for the two integrals on the right hand side of \eqref{Original_Int_3_Rewritten}. $C_1$ is again a circle with radius $R$ about $z=0$.}
 \label{Res_3}
\end{figure}

The integral \eqref{Z_D_Sinc} reads
\begin{align}
\begin{split}
 &\frac{1}{\pi}\int_0^\infty dp\,\frac{\sin^2(\xi p)}{\mu^4+p^4}\\
 &\quad=\frac{1}{2\pi}\left[\int_0^\infty dz\,\frac{1}{\mu^4+z^4}-\int_0^\infty dz\,\frac{\cos(2\xi z)}{\mu^4+z^4}\right].
 \end{split}
 \label{Original_Int_3_Rewritten}
\end{align}
The integrands of both integrals in \eqref{Original_Int_3_Rewritten} have simple poles at $z_k=\mu\,e^{i(\pi/4+\pi k/2)}$, with $k=0,1,2,3$, and the contour of integration is shown in Fig. \ref{Res_3}.
The first of the two integrals on the right hand side of \eqref{Original_Int_3_Rewritten} is readily solved with the aid of the residue theorem (again it can be shown that $\int_{C_1}dz/(\mu^4+z^4)=0$ for $R\to\infty$):
\begin{align}
\begin{split}
 &\qquad\int_0^\infty dz\,\frac{1}{\mu^4+z^4}=\frac{1}{2}\int_{-\infty}^\infty dz\,\frac{1}{\mu^4+p^4}\\
 &=\frac{1}{2}2\,\pi\,i\left[\lim_{z\to z_0}\frac{z-z_0}{(z-z_0)(z-z_1)(z-z_2)(z-z_3)}+\right.\\
 &\left.\quad+\lim_{z\to z_1}\frac{z-z_1}{(z-z_0)(z-z_1)(z-z_2)(z-z_3)}\right]\\
 &=\frac{\pi}{2\sqrt{2}\mu^3}.
 \end{split}
 \label{Original_Int_3_Final_1}
\end{align}
For the second integral it is again used that
\begin{align*}
 2\int_0^\infty dz\,\frac{\cos(2\xi z)}{\mu^4+z^4}=\int_{-\infty}^\infty\frac{e^{2i\xi z}}{\mu^4+z^4}\qquad(z\in\mathds{R}).
\end{align*}
With the integration contour shown in Fig. \ref{Res_3}, we arrive at
\begin{align*}
 &\qquad\int_{-R}^Rdz\,\frac{e^{2i\xi z}}{\mu^4+z^4}+\int_{C_1} dz\,\frac{e^{2i\xi z}}{\mu^4+z^4}\\
 &=2\,\pi\,i\left[\lim_{z\to z_0}\frac{\left(z-z_0\right)e^{2i\xi z}}{\left(z-z_0\right)\left(z-z_1\right)\left(z-z_2\right)\left(z-z_3\right)}+\right.\\
 &\left.\quad+\lim_{z\to z_1}\frac{\left(z-z_1\right)e^{2i\xi z}}{\left(z-z_0\right)\left(z-z_1\right)\left(z-z_2\right)\left(z-z_3\right)}\right]\\
 &=\frac{\pi\,e^{-\sqrt{2}\xi\mu}}{\sqrt{2}\mu^3}\left[\cos(\sqrt{2}\xi\mu)+\sin(\sqrt{2}\xi\mu)\right].
\end{align*}
As $\int_{C_1}dz e^{2i\xi z}/(\mu^4+z^4)$ tends to zero for $R\to\infty$,
\begin{align*}
 &\lim_{R\to\infty}\left|\int_0^\pi d\varphi\,\frac{iRe^{i\varphi}e^{2i\xi Re^{i\varphi}}}{\mu^4+R^4e^{4i\varphi}}\right|\\
 &\quad\leq\lim_{R\to\infty}\int_0^\pi d\varphi\,\frac{\left|iRe^{i\varphi}e^{2i\xi Re^{i\varphi}}\right|}{\left|\mu^4+R^4e^{4i\varphi}\right|}\\
 &\quad=\lim_{R\to\infty}\int_0^\pi d\varphi\frac{e^{-2\xi R\sin\varphi}}{R^3\left|\frac{\mu^4}{R^4}+e^{4i\varphi}\right|}=0,
\end{align*}
it is found that
\begin{align}
 \int_0^\infty dz\,\frac{\cos(2\xi z)}{\mu^4+z^4}=\frac{\pi\,e^{-\sqrt{2}\xi\mu}}{2\sqrt{2}\mu^3}\left[\cos(\sqrt{2}\xi\mu)+\sin(\sqrt{2}\xi\mu)\right].
 \label{Original_Int_3_Final_2}
\end{align}
The results from \eqref{Original_Int_3_Final_1} and \eqref{Original_Int_3_Final_2}, inserted into \eqref{Z_D_Sinc}, yield the expression in \eqref{Z_D_Sinc_Evaluated}.

\bibliographystyle{apsrev4-1}
\bibliography{Reference_List}

\begin{thebibliography}{24}%
\makeatletter
\providecommand \@ifxundefined [1]{%
 \@ifx{#1\undefined}
}%
\providecommand \@ifnum [1]{%
 \ifnum #1\expandafter \@firstoftwo
 \else \expandafter \@secondoftwo
 \fi
}%
\providecommand \@ifx [1]{%
 \ifx #1\expandafter \@firstoftwo
 \else \expandafter \@secondoftwo
 \fi
}%
\providecommand \natexlab [1]{#1}%
\providecommand \enquote  [1]{``#1''}%
\providecommand \bibnamefont  [1]{#1}%
\providecommand \bibfnamefont [1]{#1}%
\providecommand \citenamefont [1]{#1}%
\providecommand \href@noop [0]{\@secondoftwo}%
\providecommand \href [0]{\begingroup \@sanitize@url \@href}%
\providecommand \@href[1]{\@@startlink{#1}\@@href}%
\providecommand \@@href[1]{\endgroup#1\@@endlink}%
\providecommand \@sanitize@url [0]{\catcode `\\12\catcode `\$12\catcode
  `\&12\catcode `\#12\catcode `\^12\catcode `\_12\catcode `\%12\relax}%
\providecommand \@@startlink[1]{}%
\providecommand \@@endlink[0]{}%
\providecommand \url  [0]{\begingroup\@sanitize@url \@url }%
\providecommand \@url [1]{\endgroup\@href {#1}{\urlprefix }}%
\providecommand \urlprefix  [0]{URL }%
\providecommand \Eprint [0]{\href }%
\providecommand \doibase [0]{http://dx.doi.org/}%
\providecommand \selectlanguage [0]{\@gobble}%
\providecommand \bibinfo  [0]{\@secondoftwo}%
\providecommand \bibfield  [0]{\@secondoftwo}%
\providecommand \translation [1]{[#1]}%
\providecommand \BibitemOpen [0]{}%
\providecommand \bibitemStop [0]{}%
\providecommand \bibitemNoStop [0]{.\EOS\space}%
\providecommand \EOS [0]{\spacefactor3000\relax}%
\providecommand \BibitemShut  [1]{\csname bibitem#1\endcsname}%
\let\auto@bib@innerbib\@empty
\bibitem [{\citenamefont {Kardar}\ \emph {et~al.}(1986)\citenamefont {Kardar},
  \citenamefont {Parisi},\ and\ \citenamefont
  {Zhang}}]{Kardar_Parisi_Zhang1986}%
  \BibitemOpen
  \bibfield  {author} {\bibinfo {author} {\bibfnamefont {M.}~\bibnamefont
  {Kardar}}, \bibinfo {author} {\bibfnamefont {G.}~\bibnamefont {Parisi}}, \
  and\ \bibinfo {author} {\bibfnamefont {Y.-C.}\ \bibnamefont {Zhang}},\ }\href
  {\doibase 10.1103/PhysRevLett.56.889} {\bibfield  {journal} {\bibinfo
  {journal} {Phys. Rev. Lett.}\ }\textbf {\bibinfo {volume} {56}},\ \bibinfo
  {pages} {889} (\bibinfo {year} {1986})}\BibitemShut {NoStop}%
\bibitem [{\citenamefont {Medina}\ \emph {et~al.}(1989)\citenamefont {Medina},
  \citenamefont {Hwa}, \citenamefont {Kardar},\ and\ \citenamefont
  {Zhang}}]{Medina1989}%
  \BibitemOpen
  \bibfield  {author} {\bibinfo {author} {\bibfnamefont {E.}~\bibnamefont
  {Medina}}, \bibinfo {author} {\bibfnamefont {T.}~\bibnamefont {Hwa}},
  \bibinfo {author} {\bibfnamefont {M.}~\bibnamefont {Kardar}}, \ and\ \bibinfo
  {author} {\bibfnamefont {Y.-C.}\ \bibnamefont {Zhang}},\ }\href {\doibase
  10.1103/PhysRevA.39.3053} {\bibfield  {journal} {\bibinfo  {journal} {Phys.
  Rev. A}\ }\textbf {\bibinfo {volume} {39}},\ \bibinfo {pages} {3053}
  (\bibinfo {year} {1989})}\BibitemShut {NoStop}%
\bibitem [{\citenamefont {Forster}\ \emph {et~al.}(1977)\citenamefont
  {Forster}, \citenamefont {Nelson},\ and\ \citenamefont
  {Stephen}}]{Forster_Nelson_Stephen1977}%
  \BibitemOpen
  \bibfield  {author} {\bibinfo {author} {\bibfnamefont {D.}~\bibnamefont
  {Forster}}, \bibinfo {author} {\bibfnamefont {D.~R.}\ \bibnamefont {Nelson}},
  \ and\ \bibinfo {author} {\bibfnamefont {M.~J.}\ \bibnamefont {Stephen}},\
  }\href {\doibase 10.1103/PhysRevA.16.732} {\bibfield  {journal} {\bibinfo
  {journal} {Phys. Rev. A}\ }\textbf {\bibinfo {volume} {16}},\ \bibinfo
  {pages} {732} (\bibinfo {year} {1977})}\BibitemShut {NoStop}%
\bibitem [{\citenamefont {Janssen}\ \emph {et~al.}(1999)\citenamefont
  {Janssen}, \citenamefont {T{\"a}uber},\ and\ \citenamefont
  {Frey}}]{Janssen1999}%
  \BibitemOpen
  \bibfield  {author} {\bibinfo {author} {\bibfnamefont {H.}~\bibnamefont
  {Janssen}}, \bibinfo {author} {\bibfnamefont {U.}~\bibnamefont {T{\"a}uber}},
  \ and\ \bibinfo {author} {\bibfnamefont {E.}~\bibnamefont {Frey}},\ }\href
  {\doibase 10.1007/s100510050790} {\bibfield  {journal} {\bibinfo  {journal}
  {The European Physical Journal B - Condensed Matter and Complex Systems}\
  }\textbf {\bibinfo {volume} {9}},\ \bibinfo {pages} {491} (\bibinfo {year}
  {1999})}\BibitemShut {NoStop}%
\bibitem [{\citenamefont {Kloss}\ \emph {et~al.}(2014)\citenamefont {Kloss},
  \citenamefont {Canet}, \citenamefont {Delamotte},\ and\ \citenamefont
  {Wschebor}}]{Canet2014}%
  \BibitemOpen
  \bibfield  {author} {\bibinfo {author} {\bibfnamefont {T.}~\bibnamefont
  {Kloss}}, \bibinfo {author} {\bibfnamefont {L.}~\bibnamefont {Canet}},
  \bibinfo {author} {\bibfnamefont {B.}~\bibnamefont {Delamotte}}, \ and\
  \bibinfo {author} {\bibfnamefont {N.}~\bibnamefont {Wschebor}},\ }\href
  {\doibase 10.1103/PhysRevE.89.022108} {\bibfield  {journal} {\bibinfo
  {journal} {Phys. Rev. E}\ }\textbf {\bibinfo {volume} {89}},\ \bibinfo
  {pages} {022108} (\bibinfo {year} {2014})}\BibitemShut {NoStop}%
\bibitem [{\citenamefont {Mathey}\ \emph {et~al.}(2017)\citenamefont {Mathey},
  \citenamefont {Agoritsas}, \citenamefont {Kloss}, \citenamefont {Lecomte},\
  and\ \citenamefont {Canet}}]{Canet2017}%
  \BibitemOpen
  \bibfield  {author} {\bibinfo {author} {\bibfnamefont {S.}~\bibnamefont
  {Mathey}}, \bibinfo {author} {\bibfnamefont {E.}~\bibnamefont {Agoritsas}},
  \bibinfo {author} {\bibfnamefont {T.}~\bibnamefont {Kloss}}, \bibinfo
  {author} {\bibfnamefont {V.}~\bibnamefont {Lecomte}}, \ and\ \bibinfo
  {author} {\bibfnamefont {L.}~\bibnamefont {Canet}},\ }\href {\doibase
  10.1103/PhysRevE.95.032117} {\bibfield  {journal} {\bibinfo  {journal} {Phys.
  Rev. E}\ }\textbf {\bibinfo {volume} {95}},\ \bibinfo {pages} {032117}
  (\bibinfo {year} {2017})}\BibitemShut {NoStop}%
\bibitem [{\citenamefont {Janssen}(1976)}]{Janssen1976}%
  \BibitemOpen
  \bibfield  {author} {\bibinfo {author} {\bibfnamefont {H.-K.}\ \bibnamefont
  {Janssen}},\ }\href {\doibase 10.1007/BF01316547} {\bibfield  {journal}
  {\bibinfo  {journal} {Zeitschrift f{\"u}r Physik B Condensed Matter}\
  }\textbf {\bibinfo {volume} {23}},\ \bibinfo {pages} {377} (\bibinfo {year}
  {1976})}\BibitemShut {NoStop}%
\bibitem [{\citenamefont {De~Dominicis}(1976)}]{DeDominicis1976}%
  \BibitemOpen
  \bibfield  {author} {\bibinfo {author} {\bibfnamefont {C.}~\bibnamefont
  {De~Dominicis}},\ }\href {\doibase 10.1051/jphyscol:1976138} {\bibfield
  {journal} {\bibinfo  {journal} {{Journal de Physique Colloques}}\ }\textbf
  {\bibinfo {volume} {37}},\ \bibinfo {pages} {C1} (\bibinfo {year}
  {1976})}\BibitemShut {NoStop}%
\bibitem [{\citenamefont {Hochberg}\ \emph {et~al.}(1999)\citenamefont
  {Hochberg}, \citenamefont {Molina-Par\'{\i}s}, \citenamefont
  {P\'erez-Mercader},\ and\ \citenamefont {Visser}}]{Hochberg1999}%
  \BibitemOpen
  \bibfield  {author} {\bibinfo {author} {\bibfnamefont {D.}~\bibnamefont
  {Hochberg}}, \bibinfo {author} {\bibfnamefont {C.}~\bibnamefont
  {Molina-Par\'{\i}s}}, \bibinfo {author} {\bibfnamefont {J.}~\bibnamefont
  {P\'erez-Mercader}}, \ and\ \bibinfo {author} {\bibfnamefont
  {M.}~\bibnamefont {Visser}},\ }\href {\doibase 10.1103/PhysRevE.60.6343}
  {\bibfield  {journal} {\bibinfo  {journal} {Phys. Rev. E}\ }\textbf {\bibinfo
  {volume} {60}},\ \bibinfo {pages} {6343} (\bibinfo {year}
  {1999})}\BibitemShut {NoStop}%
\bibitem [{\citenamefont {Canet}\ \emph {et~al.}(2010)\citenamefont {Canet},
  \citenamefont {Chat\'e}, \citenamefont {Delamotte},\ and\ \citenamefont
  {Wschebor}}]{Canet2010}%
  \BibitemOpen
  \bibfield  {author} {\bibinfo {author} {\bibfnamefont {L.}~\bibnamefont
  {Canet}}, \bibinfo {author} {\bibfnamefont {H.}~\bibnamefont {Chat\'e}},
  \bibinfo {author} {\bibfnamefont {B.}~\bibnamefont {Delamotte}}, \ and\
  \bibinfo {author} {\bibfnamefont {N.}~\bibnamefont {Wschebor}},\ }\href
  {\doibase 10.1103/PhysRevLett.104.150601} {\bibfield  {journal} {\bibinfo
  {journal} {Phys. Rev. Lett.}\ }\textbf {\bibinfo {volume} {104}},\ \bibinfo
  {pages} {150601} (\bibinfo {year} {2010})}\BibitemShut {NoStop}%
\bibitem [{\citenamefont {Mu{\~{n}}oz}\ and\ \citenamefont
  {Burgett}(1989)}]{Munoz1989}%
  \BibitemOpen
  \bibfield  {author} {\bibinfo {author} {\bibfnamefont {G.}~\bibnamefont
  {Mu{\~{n}}oz}}\ and\ \bibinfo {author} {\bibfnamefont {W.~S.}\ \bibnamefont
  {Burgett}},\ }\href {\doibase 10.1007/BF01044231} {\bibfield  {journal}
  {\bibinfo  {journal} {Journal of Statistical Physics}\ }\textbf {\bibinfo
  {volume} {56}},\ \bibinfo {pages} {59} (\bibinfo {year} {1989})}\BibitemShut
  {NoStop}%
\bibitem [{\citenamefont {Hochberg}\ \emph {et~al.}(2000)\citenamefont
  {Hochberg}, \citenamefont {Molina-Par\'{\i}s}, \citenamefont
  {P\'{e}rez-Mercader},\ and\ \citenamefont {Visser}}]{Hochberg2000}%
  \BibitemOpen
  \bibfield  {author} {\bibinfo {author} {\bibfnamefont {D.}~\bibnamefont
  {Hochberg}}, \bibinfo {author} {\bibfnamefont {C.}~\bibnamefont
  {Molina-Par\'{\i}s}}, \bibinfo {author} {\bibfnamefont {J.}~\bibnamefont
  {P\'{e}rez-Mercader}}, \ and\ \bibinfo {author} {\bibfnamefont
  {M.}~\bibnamefont {Visser}},\ }\href {\doibase
  https://doi.org/10.1016/S0378-4371(99)00611-1} {\bibfield  {journal}
  {\bibinfo  {journal} {Physica A: Statistical Mechanics and its Applications}\
  }\textbf {\bibinfo {volume} {280}},\ \bibinfo {pages} {437} (\bibinfo {year}
  {2000})}\BibitemShut {NoStop}%
\bibitem [{\citenamefont {Canet}\ \emph {et~al.}(2016)\citenamefont {Canet},
  \citenamefont {Delamotte},\ and\ \citenamefont {Wschebor}}]{Canet2016}%
  \BibitemOpen
  \bibfield  {author} {\bibinfo {author} {\bibfnamefont {L.}~\bibnamefont
  {Canet}}, \bibinfo {author} {\bibfnamefont {B.}~\bibnamefont {Delamotte}}, \
  and\ \bibinfo {author} {\bibfnamefont {N.}~\bibnamefont {Wschebor}},\ }\href
  {\doibase 10.1103/PhysRevE.93.063101} {\bibfield  {journal} {\bibinfo
  {journal} {Phys. Rev. E}\ }\textbf {\bibinfo {volume} {93}},\ \bibinfo
  {pages} {063101} (\bibinfo {year} {2016})}\BibitemShut {NoStop}%
\bibitem [{\citenamefont {T{\"a}uber}(2014)}]{Taeuber2014_Book}%
  \BibitemOpen
  \bibfield  {author} {\bibinfo {author} {\bibfnamefont {U.~C.}\ \bibnamefont
  {T{\"a}uber}},\ }\href@noop {} {\emph {\bibinfo {title} {{Critical Dynamics:
  A Field Theory Approach to Equilibrium and Non-Equilibrium Scaling
  Behavior}}}}\ (\bibinfo  {publisher} {Cambridge University Press},\ \bibinfo
  {year} {2014})\BibitemShut {NoStop}%
\bibitem [{\citenamefont {Canet}\ \emph {et~al.}(2011)\citenamefont {Canet},
  \citenamefont {Chat\'e},\ and\ \citenamefont {Delamotte}}]{Canet2011_1}%
  \BibitemOpen
  \bibfield  {author} {\bibinfo {author} {\bibfnamefont {L.}~\bibnamefont
  {Canet}}, \bibinfo {author} {\bibfnamefont {H.}~\bibnamefont {Chat\'e}}, \
  and\ \bibinfo {author} {\bibfnamefont {B.}~\bibnamefont {Delamotte}},\ }\href
  {http://stacks.iop.org/1751-8121/44/i=49/a=495001} {\bibfield  {journal}
  {\bibinfo  {journal} {Journal of Physics A: Mathematical and Theoretical}\
  }\textbf {\bibinfo {volume} {44}},\ \bibinfo {pages} {495001} (\bibinfo
  {year} {2011})}\BibitemShut {NoStop}%
\bibitem [{\citenamefont {Frey}\ and\ \citenamefont
  {T{\"a}uber}(1994)}]{Taeuber1994}%
  \BibitemOpen
  \bibfield  {author} {\bibinfo {author} {\bibfnamefont {E.}~\bibnamefont
  {Frey}}\ and\ \bibinfo {author} {\bibfnamefont {U.~C.}\ \bibnamefont
  {T{\"a}uber}},\ }\href {\doibase 10.1103/PhysRevE.50.1024} {\bibfield
  {journal} {\bibinfo  {journal} {Phys. Rev. E}\ }\textbf {\bibinfo {volume}
  {50}},\ \bibinfo {pages} {1024} (\bibinfo {year} {1994})}\BibitemShut
  {NoStop}%
\bibitem [{\citenamefont {Zinn-Justin}(2002)}]{Zinn_Justin2002}%
  \BibitemOpen
  \bibfield  {author} {\bibinfo {author} {\bibfnamefont {J.}~\bibnamefont
  {Zinn-Justin}},\ }\href@noop {} {\emph {\bibinfo {title} {{Quantum Field
  Theory and Critical Phenomena}}}},\ International Series of Monographs on
  Physics\ (\bibinfo  {publisher} {Oxford University Press},\ \bibinfo {year}
  {2002})\BibitemShut {NoStop}%
\bibitem [{\citenamefont {Zinn-Justin}(2007)}]{Zinn_Justin2007}%
  \BibitemOpen
  \bibfield  {author} {\bibinfo {author} {\bibfnamefont {J.}~\bibnamefont
  {Zinn-Justin}},\ }\href@noop {} {\emph {\bibinfo {title} {{Phase Transitions
  and Renormalization Group}}}}\ (\bibinfo  {publisher} {Oxford University
  Press},\ \bibinfo {year} {2007})\BibitemShut {NoStop}%
\bibitem [{\citenamefont {Amit}(1984)}]{Amit1984}%
  \BibitemOpen
  \bibfield  {author} {\bibinfo {author} {\bibfnamefont {D.}~\bibnamefont
  {Amit}},\ }\href {https://books.google.de/books?id=gEVVcAAACAAJ} {\emph
  {\bibinfo {title} {Field Theory, the Renormalization Group, and Critical
  Phenomena}}}\ (\bibinfo  {publisher} {World Scientific},\ \bibinfo {year}
  {1984})\BibitemShut {NoStop}%
\bibitem [{\citenamefont {Mussardo}(2010)}]{Mussardo2010}%
  \BibitemOpen
  \bibfield  {author} {\bibinfo {author} {\bibfnamefont {G.}~\bibnamefont
  {Mussardo}},\ }\href@noop {} {\emph {\bibinfo {title} {{Statistical Field
  Theory -- An Introduction to Exactly Solved Models in Statistical
  Physics}}}}\ (\bibinfo  {publisher} {Oxford University Press},\ \bibinfo
  {year} {2010})\BibitemShut {NoStop}%
\bibitem [{\citenamefont {Family}\ and\ \citenamefont
  {Vicsek}(1985)}]{Family1985}%
  \BibitemOpen
  \bibfield  {author} {\bibinfo {author} {\bibfnamefont {F.}~\bibnamefont
  {Family}}\ and\ \bibinfo {author} {\bibfnamefont {T.}~\bibnamefont
  {Vicsek}},\ }\href {\doibase 10.1088/0305-4470/18/2/005} {\bibfield
  {journal} {\bibinfo  {journal} {Journal Physics D: Applied Physics}\ }\textbf
  {\bibinfo {volume} {18}} (\bibinfo {year} {1985}),\
  10.1088/0305-4470/18/2/005}\BibitemShut {NoStop}%
\bibitem [{\citenamefont {Kim}\ and\ \citenamefont
  {Kosterlitz}(1989)}]{Kosterlitz1989}%
  \BibitemOpen
  \bibfield  {author} {\bibinfo {author} {\bibfnamefont {J.~M.}\ \bibnamefont
  {Kim}}\ and\ \bibinfo {author} {\bibfnamefont {J.~M.}\ \bibnamefont
  {Kosterlitz}},\ }\href {\doibase 10.1103/PhysRevLett.62.2289} {\bibfield
  {journal} {\bibinfo  {journal} {Phys. Rev. Lett.}\ }\textbf {\bibinfo
  {volume} {62}},\ \bibinfo {pages} {2289} (\bibinfo {year}
  {1989})}\BibitemShut {NoStop}%
\bibitem [{\citenamefont {Wilson}(1975)}]{Wilson1975}%
  \BibitemOpen
  \bibfield  {author} {\bibinfo {author} {\bibfnamefont {K.~G.}\ \bibnamefont
  {Wilson}},\ }\href {\doibase 10.1103/RevModPhys.47.773} {\bibfield  {journal}
  {\bibinfo  {journal} {Rev. Mod. Phys.}\ }\textbf {\bibinfo {volume} {47}},\
  \bibinfo {pages} {773} (\bibinfo {year} {1975})}\BibitemShut {NoStop}%
\bibitem [{\citenamefont {Halpin-Healy}\ and\ \citenamefont
  {Zhang}(1995)}]{HalpinHealy1995}%
  \BibitemOpen
  \bibfield  {author} {\bibinfo {author} {\bibfnamefont {T.}~\bibnamefont
  {Halpin-Healy}}\ and\ \bibinfo {author} {\bibfnamefont {Y.-C.}\ \bibnamefont
  {Zhang}},\ }\href {\doibase http://dx.doi.org/10.1016/0370-1573(94)00087-J}
  {\bibfield  {journal} {\bibinfo  {journal} {Physics Reports}\ }\textbf
  {\bibinfo {volume} {254}},\ \bibinfo {pages} {215 } (\bibinfo {year}
  {1995})}\BibitemShut {NoStop}%
\end{thebibliography}%

\end{document}